\documentclass[11pt]{article}
\usepackage[utf8]{inputenc}
\usepackage{graphicx}
\usepackage{float}
\usepackage{subfigure}
\usepackage{caption}
\usepackage{mathtools}
\usepackage{times}
\usepackage{url}

\usepackage{indentfirst}

\usepackage{hyperref}

\usepackage[none]{hyphenat}

\usepackage{ragged2e}

\usepackage[hyperref=false,backend=biber,sorting=none,backref=true]{biblatex}
\title{Investigation of the Relationship Between Localization Accuracy and Sensor Array}
\author{Yiyang Li  \\  y.li04@umcg.nl }

\date{}

\begin{document}

\maketitle

\justifying  
\begin{abstract}

     \noindent
     \sloppy{}
    {The magnetic localization method has been widely studied, which is mainly based on the accurate mapping of the magnetic field generated by magnetic sources. Many factors affect localization accuracy in the experiment. 
    Therefore, this paper tends to study the relationship between localization accuracy and sensor array with different experiments. This system uses a small magnet as the magnetic source, and the mathematical model of the magnetic positioning system is established based on the magnetic dipole model to estimate the magnetic field. 
    The Levenberg-Marquardt algorithm was used to construct a magnetic positioning objective function for comparison experiments. Experimental results show:When the sensor is evenly distributed around the magnet, the positioning accuracy is higher than other layout of the sensor array, the  average localization error is 0.47mm and the average orientation error is 0.92 degree. }

    
\end{abstract}


\sloppy{}
\section{Introduction}

Magnetic microrobots are a promising and significant tool in the medical field, they are used for micromanipulation and minimally invasive interventions on biological cells, such as kidney stone removal, intravascular drug delivery, and thrombus removal\cite{9226520, 8845324,9278687}. 
In these applications, it is very important to accurately obtain the position of the magnetic microrobots\cite{9226520}, because it can help doctors to obtain the position and size of the patient’s lesion tissue more accurately and comprehensively, thereby completing more efficient and reliable treatment\cite{8664916,mateen2017localization}. And is also used for feedback control magnetic micro robot \cite{mateen2017localization}.

Many scholars and experts have done a lot of research on the positioning of micro robots, and there are many positioning methods. Different positioning methods have their own advantages and disadvantages. The positioning technology based on radio frequency RF signal is cheap and easy to operate, but the positioning accuracy is too low\cite{brusey2009effective,1307283,4374110}.
Based on the positioning methods of medical imaging technology, these technologies require some huge equipments, which are expensive; and the operation is complicated and the positioning accuracy is also limited\cite{8363549,8721095,7493234}. 
Based on ultrasonic positioning technology, ultrasonic signals are easily interfered by reflective objects in the surrounding environment. When the ultrasonic signal attenuates, it will affect the measurement range, and these factors will affect the positioning accuracy\cite{8492639,8778619,9166535}. Based on the positioning technology of machine vision, this method has high positioning accuracy and relatively comprehensive information, but it needs to process a large amount of data, which requires high processing capabilities of the processor, which leads to increased investment costs\cite{8496860,8492837,9084655}.
The positioning method based on the static magnetic field uses permanent magnets as the magnetic source and tracking target. The magnetic field intensity generated by the magnet in space is related to the distance from the point to the magnet and the orientation of the magnet.Use the magnetic sensor to detect the magnetic field intensity, and calculate the position and orientation information of the magnet inversely according to a specific positioning algorithm\cite{Hu2010AC3,6782730,7458794}. The magnetic intensity of the magnet are changes as a function of the static magnetic field and is not affected by the human body. And the magnetic microrobot does not need power source energy, no line-of-sight restriction and a wireless device brings more convenience\cite{7989138,5262602,8624518}.

A great deal of researchers have proposed a large number of localization approach for the magnetic microrobot's  based on the model of the magnetic field positioning \cite{5741733,8234258,5949026}. Generally, we utilize magnetic sensors to measure the magnetic intensity of the magnetic microrobot, the position and orientation of the magnetic microrobot can be estimated via some optimization algorithms\cite{8519764,8603829}.
For instance, Chao et al. proposed to use the magnetic dipole model for the positioning of the wireless capsule endoscope, and used a Levenberg-Marquardt(LM) nonlinear algorithm to infer the position and orientation of the magnet step by step according to the sampled magnetic field, which get the results with the localization error smaller than 10mm\cite{1616154}. Subsequently, a linear analytical algorithm was proposed to simplify the algorithm, which with the average localization error of magnetic microrobot is 1.2mm\cite{1616154,4649596}. Then, Chao et al. proposed  the linear algorithm and the Levenberg-Marquardt(LM) algorithm based on a cubic sensor array, with the average localization error of 1.8 mm and 1.54 degree.
 At present, there are a great deal of innovative methods proposed widely to improve the positioning accuracy based on the magnet positioning algorithms \cite{5753941,7165601,7088545,7156055,5170236}, and these positioning methods have become mature. In fact, how to obtain accurate data is also one of the important steps affecting the experiment. However, few researchers informed us how to collect data effectively. Shuang et al.  only proposed methods to optimize sensor layout\cite{8013792,Song2017DesignAO}. Hence, this paper tends to investigate relationship between localization accuracy and sensor array.

In this paper, a mathematical model of the magnetic positioning system based on the magnetic dipole model is established. We use Levenberg-Marquardt(LM) algorithm \cite{5420750,7593319} to construct the magnetic positioning objective function. According to the output of the magnetic sensors, an initial value of the position and direction parameters of the magnet is calculated by the linear algorithm, the optimal solutions of the magnet are derived by the nonlinear iterative algorithm. We use multiple sets of comparative experiments to investigate relationship between localization accuracy and sensor array.

\section{Mathmatic Model and Mininization Alogorithm}

\subsection{Mathematic Model}

Currently, the magnetic dipole model is widely used in the positioning and state determination of target objects\cite{5741733,8234258}. When the size of the permanent magnet is much smaller than the distance between the detection point and the permanent magnet, the permanent magnet can be equivalent to a magnetic dipole. 
 
The cylindrical permanent magnet with length $L$, and diameter $r$ in the experiment.Compare to the distance between the magnetic sensors and the magnet, the size and the magnet is much smaller, therefore, we can use the magnetic dipole model and define a fixed coordinate system as shown in Figure 1. Under the coordinate system, let $\text P$ be the vector that represents a  spatial point $\text l=\left(x_{l}, y_{l}, z_{l}\right)^{T}$ with  respect to the magnet position $ \left(a,b,c\right)^{T}$  as shown in Figure 1.


\begin{figure}	
\center
\includegraphics[width=7cm,height=6cm]{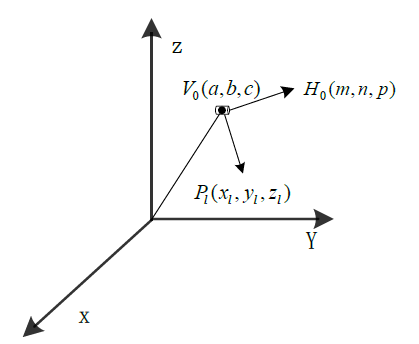}
\caption{Coordinate System for Magnet's Localization.$V_{0}$ is the center of the magnet,  
 $\left(x_{l}, y_{l}, z_{l}\right)^{T}$  is the position of the l-th  sensor, $H_{0}$ is a unit vector of the magnet,  $ \left(m,n,p\right)^{T}$ representing the magnet’s orientation from south pole to north pole, $P_{l}$  denotes the vector from  $(a,b,c)$  to $\left(x_{l}, y_{l}, z_{l}\right)$. }  
\end{figure}  

In order to study the relationship between sensor layout and positioning accuracy, we use the following two different sensor layout methods to conduct experiments:
the first sensor layout mode adopts a unified 2*n (n=8) symmetrical mode as shown in Figure 2.(a); the second sensor layout mode adopts a unified 4*m (m=5) symmetrical mode as shown in Figure 2.(b).

\begin{figure}
\centering
\subfigure[2*n]{
\begin{minipage}[t]{0.45\linewidth}
\includegraphics[width=5cm,height=5cm]{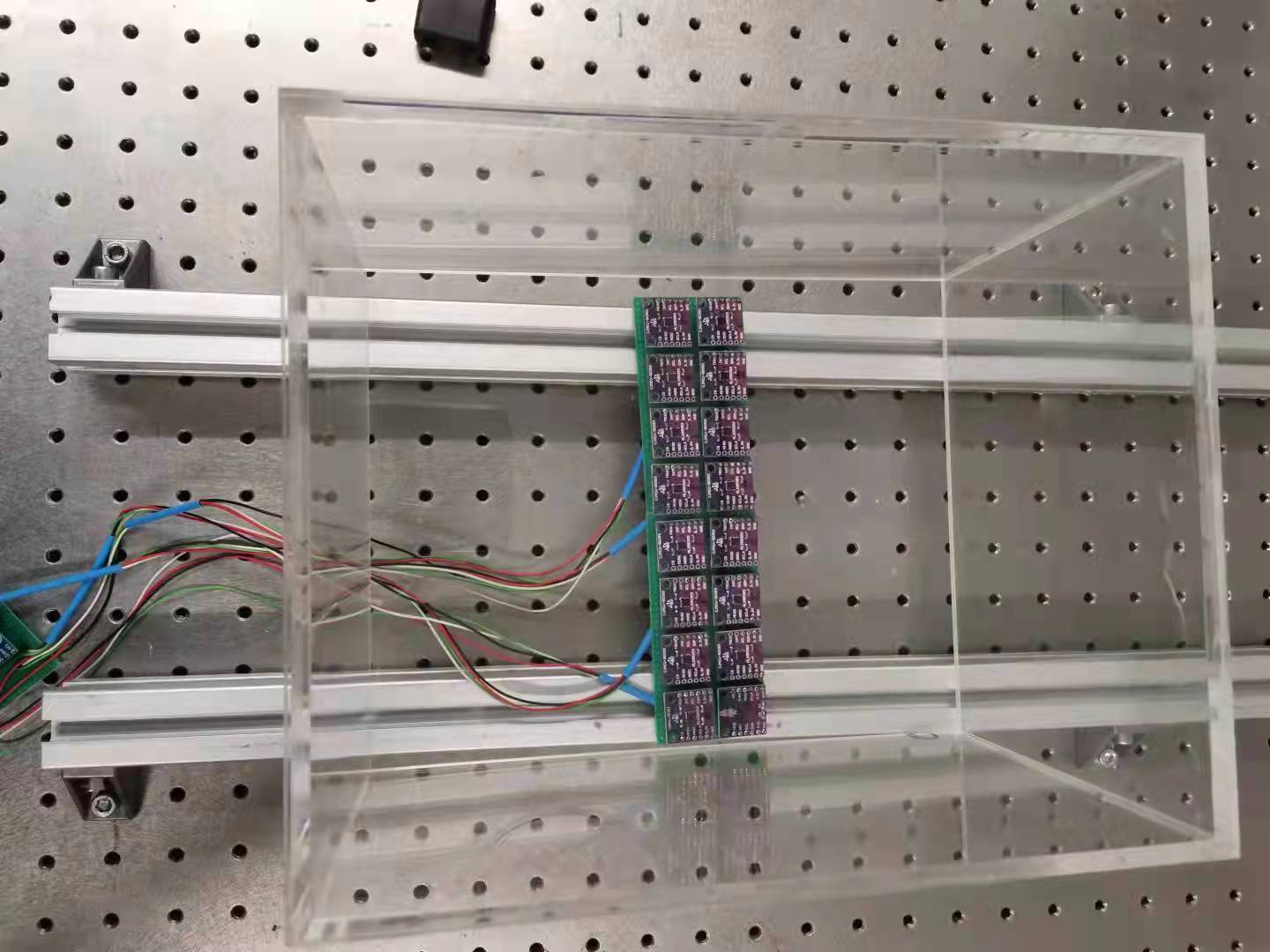}
\end{minipage}%
}%
\subfigure[4*m]{
\begin{minipage}[t]{0.45\linewidth}

\includegraphics[width=5cm,height=5cm]{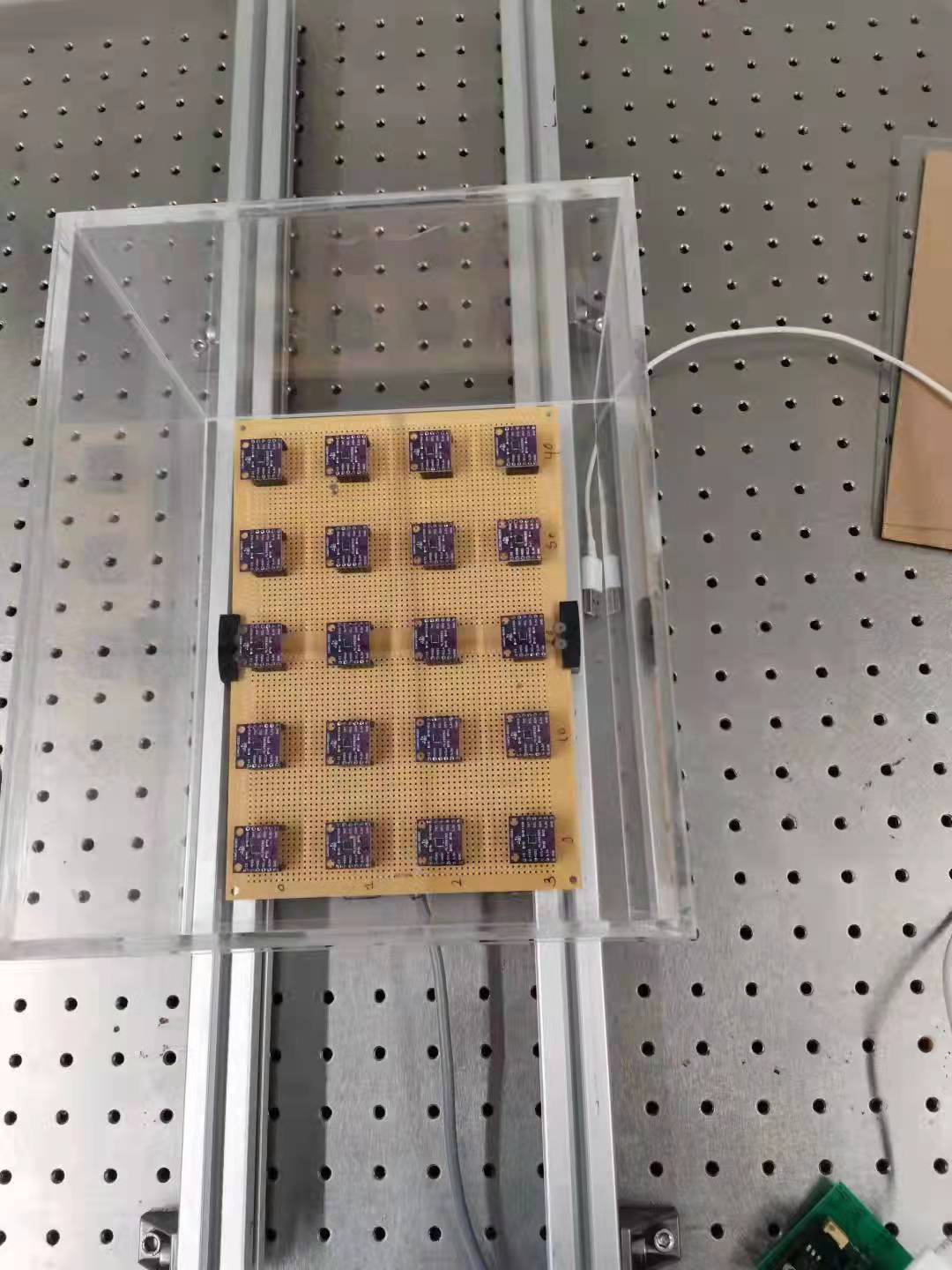}
\end{minipage}%
}%

\caption{All sensors are arranged in the same X-Y plane in those two different layout sensor array. (a)Planar array was based on 16 digital magnetic sensors and a single  PCB.  16 sensors were evenly organized in a horizontal plane(a single PCB with the dimension of  5cm X 20cm). The distance between adjacent sensors is 2 mm.(b)Planar array was based on 20 digital magnetic sensors and a single  PCB. 20 sensors were evenly organized in a horizontal plane(a single PCB with the dimension of  16cm X 25cm). The distance between adjacent sensors is 30 mm.
}
\end{figure}

Assume that there are $\text N$ sensors, with l-th sensor located at $\left(x_{l}, y_{l}, z_{l}\right)^{T}$, $\text1 \leq l \leq N$. The magnetic flux density at the l-th sensor location can be represented: 

\begin{equation}
B_{l}=B_{l x} i+B_{l y} j+B_{l z} k=B_{T}\left(\frac{3\left(H_{0} \cdot P_{l}\right) P_{l}}{R_{l}^{5} }-\frac{H_{0}}{R_{l}^{3}}\right), B_{T}=\frac{\mu_{r} \mu_{0} \pi r^{2} L M_{0}}{4 \pi}
\end{equation}

Where $\text R_{l}=\sqrt{\left(x_{l}-a\right)^{2}+\left(y_{l}-b\right)^{2}+\left(z_{l}-c\right)^{2}} $, $\mu_{r}$ is the relative permeability of the medium(in the air, $\mu_{r} \approx 1$), $\mu_{0}$ is the air magnetic permeability($u_{0}=4 \pi \cdot 10^{-7} \mathrm{~T} \cdot \mathrm{m} / \mathrm{A}$ ), and $\mathbf{H}_{0}=(m, n, p)^{\mathrm{T}}$ is the normalized vector representing the direction of the magnet’s magnetism.

Tri-axis magnetic sensors can be used to measure the magnetic field intensity around a magnet. Using the measured data and the mathematical model (1), the magnet’s position and orientation parameters can be computed with an appropriate nonlinear optimization algorithm. Extending (1), we have

\begin{equation}
B_{l x}=B_{T}\left\{\frac{3\left[m\left(x_{l}-a\right)+n\left(y_{l}-b\right)+p\left(z_{l}-c\right)\right] \cdot\left(x_{l}-a\right)}{R_{l}^{5}}-\frac{m}{R_{l}^{3}}\right\}
\end{equation}

\begin{equation}
B_{l y}=B_{T}\left\{\frac{3\left[m\left(x_{l}-a\right)+n\left(y_{l}-b\right)+p\left(z_{l}-c\right)\right] \cdot\left(y_{l}-b\right)}{R_{l}^{5}}-\frac{n}{R_{l}^{3}}\right\}
\end{equation}

\begin{equation}
B_{l z}=B T\left\{\frac{3\left[m\left(x_{l}-a\right)+n\left(y_{l}-b\right)+p\left(z_{l}-c\right)\right] \cdot\left(z_{l}-c\right)}{R_{l}^{5}}-\frac{p}{R_{l}^{3}}\right\}
\end{equation}

where $B_{l x}$, $B_{l y}$, and $B_{l z}$ are the three components of the magnetic field intensity along X, Y, and Z axes.
In the proposed endoscopic system, the flux intensities $B_{l x}$, $B_{l y}$, and $B_{l z}$ are measured using the l-th
magnetic sensor located around position $\left(x_{l}, y_{l}, z_{l}\right)^{\mathrm{T}}$, which is known in advance. Note that there are six unknown parameters $(a, b, c, m, n, p)$. The flux intensity is invariant to the rotation of the magnet along its major axis. Hence the magnet’s orientation $\mathbf{H}_{0}$ is in two dimensions. Therefore we added the following constraint for $(m, n, p)^{\mathrm{T}}$:

\begin{equation}
m^{2}+n^{2}+p^{2}=1
\end{equation}

\subsection{Localization Algorithms}
Supposing $\left(B_{l x}, B_{l y}, B_{l z}\right)^{T}$ and $\left(B_{l x}^{\prime}, B_{l y}^{\prime}, B_{l z}^{\prime}\right)^{T}$ are the theoretical values and measured values by the sensors, the error objective function can be defined as:
\begin{equation}
E_{x}=\sum_{l=1}^{N}\left[B_{l x}^{\prime}-B_{l x}\right]^{2}
\end{equation}

\begin{equation}
E_{y}=\sum_{l=1}^{N}\left[B_{l y}^{\prime}-B_{l y}\right]^{2}
\end{equation}

\begin{equation}
E_{z}=\sum_{l=1}^{N}\left[B_{l z}^{\prime}-B_{l z}\right]^{2}
\end{equation}

where $B_{l x}$, $B_{l x}$, and $B_{l x}$ are defined by $(2)\sim(4)$, and  $B_{l x}^{\prime}$,$B_{l y}^{\prime}$, and $B_{l z}^{\prime}$ are the three measured data of the l-th 3-axis magnetic sensor. The total error is the summation of above three errors:$E=E_{x}+E_{y}+E_{z}$. We need to find optimal parameters (a, b, c, m, n, p) to minimize this error E. This is the least square error problem, which can be solved by Levenberg-Marquardt minimization algorithms. 

\begin{equation}
\begin{array}{l}
\underset{a, b, c, m, n, p}{\arg } \min E=E_{x}+E_{y}+E_{z} \\
\text { s.t. } m^{2}+n^{2}+p^{2}=1
\end{array}
\end{equation}

To evaluate the performance of an algorithm, we define two parameters, localization error and orientation error as follows:
\begin{equation}
E_{p}=\sqrt{\left(a_{c}-a_{t}\right)^{2}+\left(b_{c}-b_{t}\right)^{2}+\left(c_{c}-c_{t}\right)^{2}}
\end{equation}

\begin{equation}
E_{o}=\sqrt{\left(m_{c}-m_{t}\right)^{2}+\left(n_{c}-n_{t}\right)^{2}+\left(p_{c}-p_{t}\right)^{2}}
\end{equation}

\begin{equation}
\theta=\arccos \left|\left(\frac{\left(m_{c}, n_{c}, p_{c}\right) \cdot\left(m_{t}, n_{t}, p_{t}\right)}{\left\|\left(m_{c}, n_{c}, p_{c}\right)\right\| \times\left\|\left(m_{t}, n_{t}, p_{t}\right)\right\|}\right)\right|
\end{equation}

where $\left(a_{c}, b_{c}, c_{c}, m_{c}, n_{c}, p_{c}\right)$ represent the calculated location and orientation parameters, and $\left(a_{t}, b_{t}, c_{t}, m_{t}, n_{t}, p_{t}\right)$ represent the true location and orientation parameters of the magnet.


\section{Experiments and discuss}
Magnetic sensors play an important role in improving the performance of the magnetic localization system. To guarantee the performance, it is essential for the sensors to be of high sensitivity, wide range and strong anti-interference ability.

Magnetic sensor MLX90393 was adopted in the experiment. Magnetic sensor MLX90393  provides 44,000uT maximum full-scale resolution with a resolution rate of 0.161uT.

When we use Hall-effect sensors to collect data, some sensors will display the data first. These data are not displayed strictly according to the location of the sensor. therefore,  we need to let the program run for one cycle at first and remove the data that is less than one cycle for  experimentation. 
During the sampling process, the data will be affected by various clutters. For this, we use the method of adding filtering to denoise.

Figure 3.(a) and figure 3.(b) is the corresponding magnetic field intensity distribution diagram obtained by measurement with or without filtering, figure 3.(c) is the difference of magnetic field intensity under noise interference.

\begin{figure}
\centering
\subfigure[without filter]{
\begin{minipage}[t]{0.25\linewidth}
\includegraphics[width=1.5in]{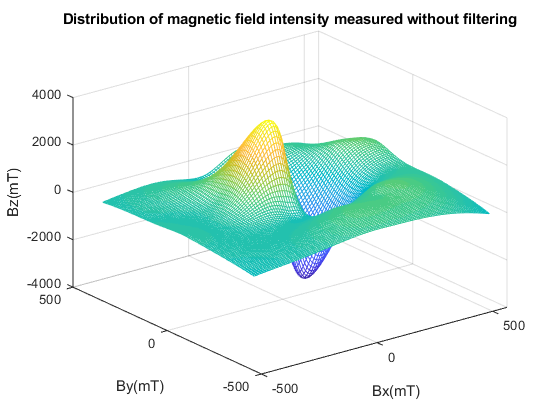}
\end{minipage}%
}%
\subfigure[with filter]{
\begin{minipage}[t]{0.25\linewidth}

\includegraphics[width=1.5in]{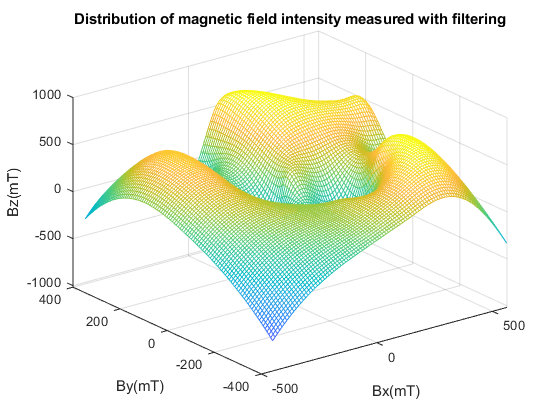}
\end{minipage}%
}%
\subfigure[the difference]{
\begin{minipage}[t]{0.25\linewidth}
\includegraphics[width=1.5in]{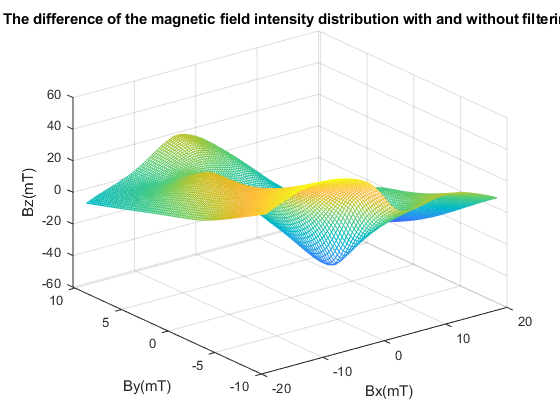}
\end{minipage}
}%

\caption{ (a)The data distribution of magnetic field intensity when filtering is not added.(b) The data distribution of magnetic field intensity with filtering. (c) The error  data distribution of magnetic field intensity under noise interference}
\end{figure}

Compare the noise error calculated when filtering is used for denoising in the experiment as figure 3. Compared to without filter to data collection, average error of magnetic field intensity along x-,y- and z axis is 0.13mT, 1.735mT, 1.153mT, respectively.

We conduct 100 experiments and take data every 4 groups  with both using filtering and without filtering. A total of 25 groups of data are used. Levenberg-Marquardt algorithm[] was used to calculate the position and orientation of the magnet.

Figure 4.(a) and figure 4.(b) describe the position distribution of the magnet in the space coordinate system with or without filtering. Figure 4.(c) and figure 4.(d)  show the position distribution in the plane coordinate system with or without filtering. The experiment in Figure 4 was carried out under the sensor layout of Figure 2.(a)(2*n).


\begin{figure}[htbp]
\centering
\subfigure[without filter]{
\includegraphics[width=5.5cm]{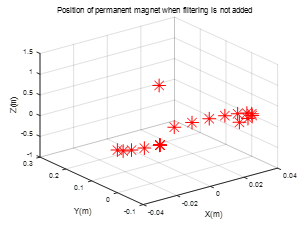}
}
\quad
\subfigure[with filter]{
\includegraphics[width=5.5cm]{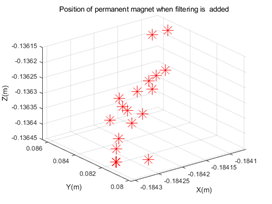}
}

\quad
\subfigure[without filter]{
\includegraphics[width=5.5cm]{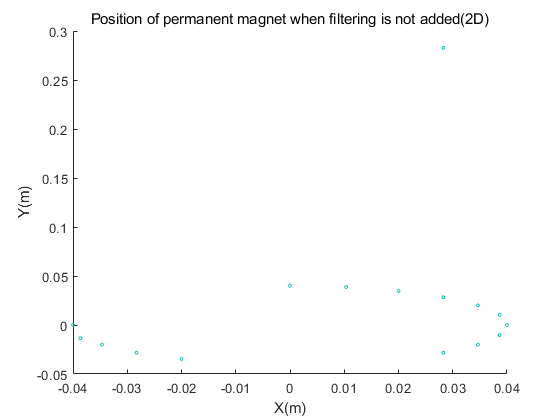}
}
\quad
\subfigure[with filter]{
\includegraphics[width=5.5cm]{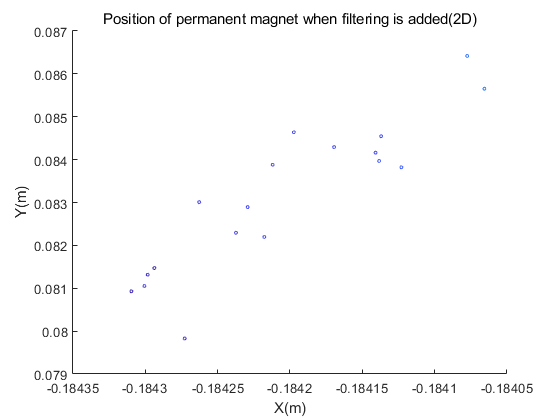}
}
\caption{ Position of permanent magnet when filtering is added or not in the space/plane coordinate system under the sensor layout in Figure 2.(a)}
\end{figure}


Figure 5.(a) and figure 5.(b) describe the position distribution of the magnet in the space coordinate system with or without filtering. Figure 5.(c) and figure 5.(d)  show the position distribution in the plane coordinate system with or without filtering. The experiment in Figure 5 was carried out under the sensor layout of Figure 2.(b)(4*m).

\begin{figure}[htbp]
\centering
\subfigure[without filter]{
\includegraphics[width=5.5cm]{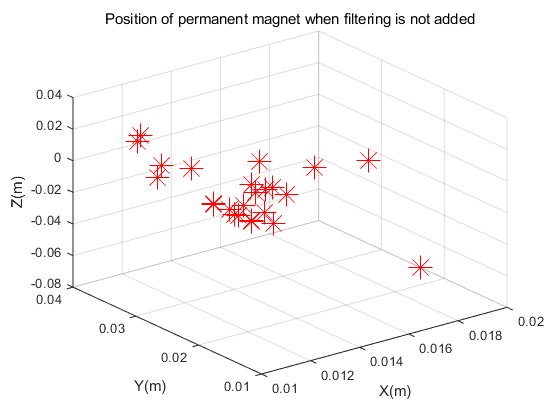}
}
\quad
\subfigure[with filter]{
\includegraphics[width=5.5cm]{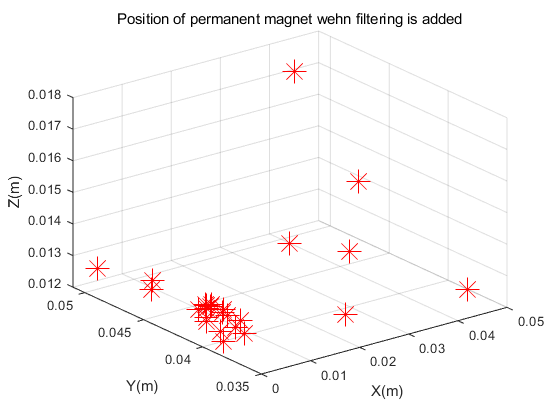}
}

\quad
\subfigure[without filter]{
\includegraphics[width=5.5cm]{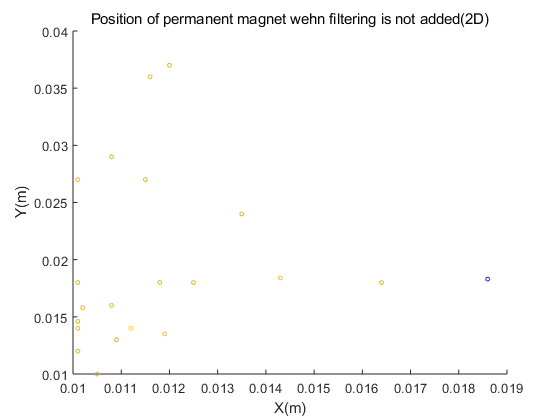}
}
\quad
\subfigure[with filter]{
\includegraphics[width=5.5cm]{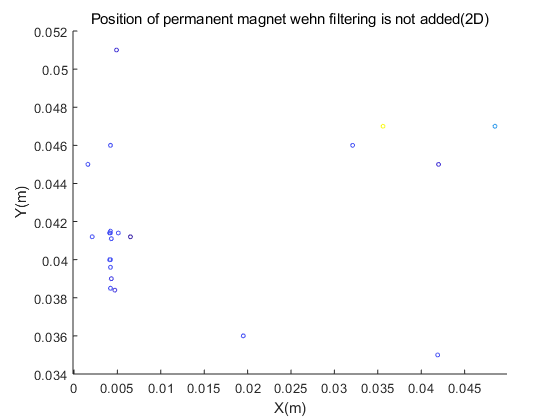}
}
\caption{ Position of permanent magnet when filtering is added or not in the space/plane coordinate system under the sensor layout in Figure 2.(b)}
\end{figure}

Figure 5 shows the position of the magnet in the sensor layout of Figure 3(4*m) in this paper. It can be seen from the figure that after adding filtering for noise reduction, the positions we get are relatively concentrated. And the experimental results obtained without filtering are obviously scattered. However, the results of the experiment in Figure 4 with the sensor layout of Figure 2.(a) in this article are completely opposite. According to theory, adding filtering will reduce noise interference, and the calculated position should be more accurate than the data without filtering, but the experimental results show that the calculation error is smaller without filtering. We guess that this may be due to the fact that the direct spacing of the sensors in Figure 2.(a) is too small, which may cause larger errors. For this reason, we have compared the position errors for experimental analysis.

Use the Levenberg -Marquardt algorithm algorithm to calculate the position of the magnet. When using the sensor layout of Figure 2.(a)(2*n), the average positioning error obtained by adding filtering and denoising at this time is 14.97 mm, and the average positioning error obtained without filtering is 17.85 mm. The same method, Using the sensor layout of Figure 2.(b)(4*m), the average positioning error obtained by adding filtering and denoising at this time is 3.32 mm, and the average positioning error obtained without filtering is 12.06 mm.
Compared with the results, we conclude that add some filters denoising method is used for positioning, the positioning error obtained at the time is much smaller.


\subsection{Experiment-1}

In order to verify the influence of the number of sensors on the positioning accuracy, we use 6-, 8-, 12-, 16-, and 20-sensor to conduct experiments in the same environment. 

Two different sensor layout methods are used to compare and contrast experiments. The first sensor layout mode adopts a unified 2*n (n=3,4,6,8) symmetrical mode,
and the second sensor layout mode adopts a unified 4*m (m=2,3,4,5) symmetrical mode.

Use formula (10), (11), (12) to calculate position error and orientation of the magnet's error. Figure 9 is the case of the first sensor layout(2*n, n=3,4,6,8), using different numbers of sensors to conduct experiments to compare positioning accuracy. Figure 10 is the case of the second sensor layout(2*m, m=2,3,4,5), using different numbers of sensors to conduct experiments to compare positioning accuracy. 

\begin{figure}[htbp]
\centering
\subfigure[2*n]{
\begin{minipage}[t]{0.45\linewidth}
\includegraphics[width=2.5in]{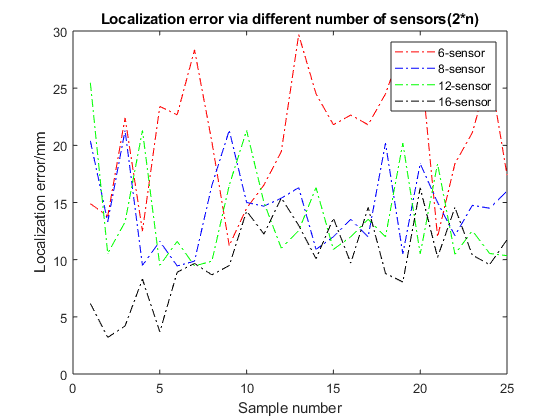}
\end{minipage}%
}%
\subfigure[4*m]{
\begin{minipage}[t]{0.45\linewidth}

\includegraphics[width=2.5in]{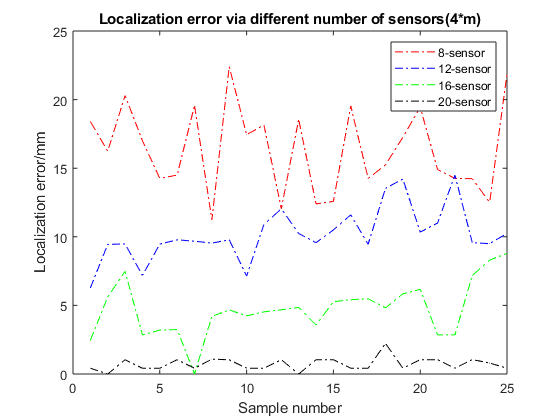}
\end{minipage}%
}%

\caption{ Localization error with different sensor layout. (a) with layout 2*n(figure 2.(a)). (b) with layout 2*n(figure 2.(b))
}
\end{figure}

\begin{figure}[htbp]
\centering
\subfigure[2*n]{
\begin{minipage}[t]{0.45\linewidth}
\includegraphics[width=2.5in]{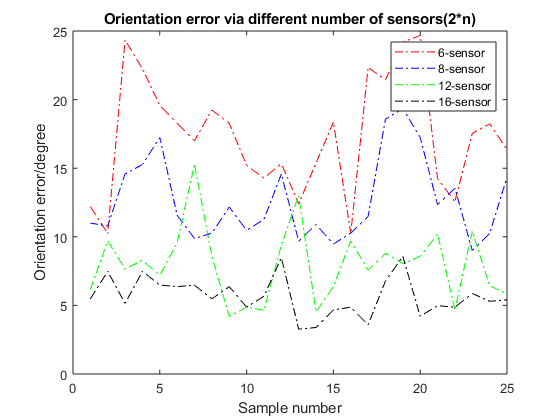}
\end{minipage}%
}%
\subfigure[4*m]{
\begin{minipage}[t]{0.45\linewidth}

\includegraphics[width=2.5in]{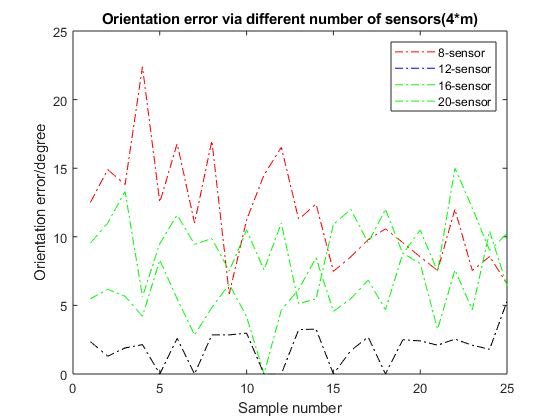}
\end{minipage}%
}%

\caption{ Orientation error with different sensor layout. (a) with layout 2*n(figure 2.(a)). (b) with layout 2*n(figure 2.(b)).
}
\end{figure}

In the case of the first sensor layout(2*n, n=3,4,6,8), the average localization error within 23.44mm, 15.17mm, 13.8mm, and 10mm  
for 6-, 8-, 12-, and 16-sensor respectively; in the case of the second sensor layout(4*m, m=2,3,4,5), the average localization error within 16.34mm, 10.2mm, 4.74mm, and 0.72mm for 8-, 12-, 16-, and 20-sensor respectively(as shown in the figure 6). Comparing the experimental results, we found that as the number of sensors increases, the positioning error of the magnet decreases. In the process of using 16 sensors to 20 sensors, the positioning error is reduced by 4.02 mm. When the number of sensors reaches a certain level, the positioning error does not change. At the same time, too many sensors will affect the laboratory execution time. Considering comprehensively, this article finally selects 20 sensors for experiments.
In the case of the first sensor layout(2*n, n=3,4,6,8), the average orientation error within 18.03 degree, 12.62 degree, 8degree, and 5.7 degree for 6-, 8-, 12-, and 16-sensor respectively; in the case of the second sensor layout(4*m, m=2,3,4,5), the average orientation error within 11.58 degree, 9.79 degree, 5.75 degree, and 1.35 degree for 8-, 12-, 16-, and 20-sensor respectively(as shown in the figure 7). 
In experiments with the same number of sensors(8-, 12-, and 16-sensor), we found that the positioning error of using sensor layout 2(figrue 2.(b)) is much smaller than that of layout 1(figrue 2.(a)). Therefore, the follow-up experiments in this article are all carried out in the case of sensor layout 2(figrue 2.(b)).



\subsection{Experiment-2}


In order to verify the use range of Levenberg-Marquardt(LM) algorithm, we placed magnets at different heights and different horizontal distances from the sensor for comparison experiments.

The prerequisite for using the magnetic dipole model is that the distance between the magnet and the magnetic sensor must be greater than the size of the magnet itself. The magnet used in this experiment is a cylinder with a length of 2 mm and a bottom radius of 1 mm. The size of the Hall-effect sensor is 26mm*26.3 mm in the experiment. Therefore, we choose the change of height and distance to be 50mm increments for each experiment.

Figures 8 shows the positioning accuracy and orientation accuracy when the magnet is at different Vertical heights.

\begin{figure}[htbp]
\centering
\subfigure[Localization error]{
\begin{minipage}[t]{0.45\linewidth}
\includegraphics[width=2.5in]{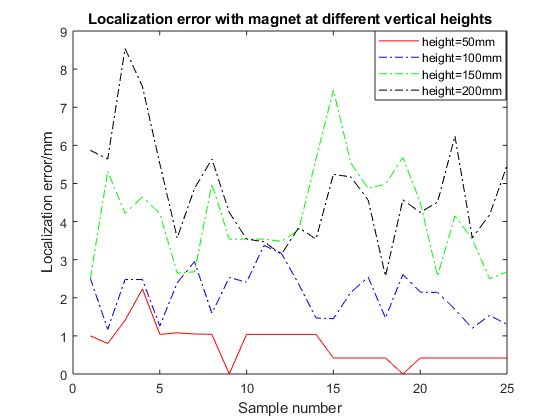}
\end{minipage}%
}%
\subfigure[Orientation error]{
\begin{minipage}[t]{0.45\linewidth}

\includegraphics[width=2.5in]{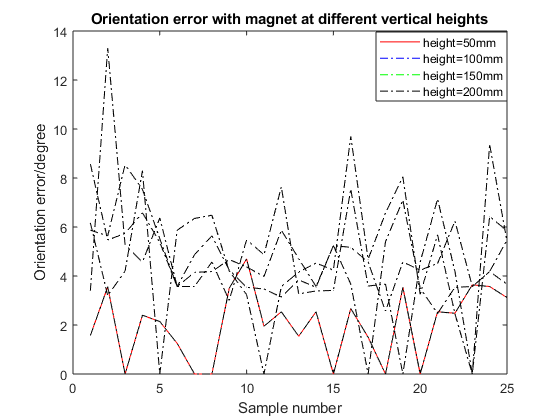}
\end{minipage}%
}%

\caption{ The localization error and orientation error of magnet at different vertical height}
\end{figure}

When the magnet is at a height of 50 mm from the sensor array, the average positioning error at this time is 0.762 mm, and the average orientation error is 2.1 degrees.
When the magnet is at a height of 100 mm from the sensor array, the average positioning error at this time is 2.1 mm, and the average orientation error is 4.57 degrees.
When the magnet is at a height of 150 mm from the sensor array, the average positioning error at this time is 4.13 mm, and the average orientation error is 4.7 degrees.
When the magnet is at a height of 200 mm from the sensor array, the average positioning error at this time is 4.77 mm, and the average orientation error is 4.8 degrees.
It can be obtained from the experimental results that when the height of the magnet is between 0 and 200 mm, the experimental results of positioning error and orientation error are not much different.

Figures 9 shows the positioning accuracy and orientation accuracy when the magnet is at different horizontal distance.

\begin{figure}[htbp]
\centering
\subfigure[Localization error]{
\begin{minipage}[t]{0.45\linewidth}
\includegraphics[width=2.5in]{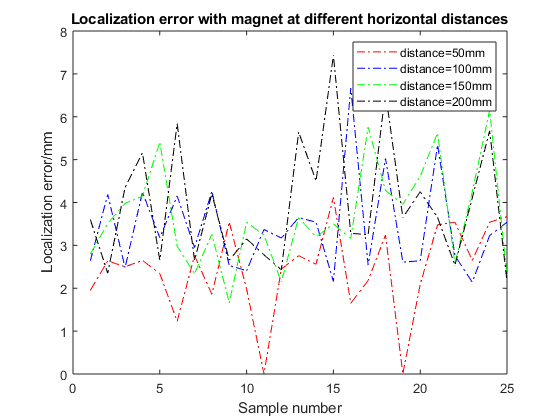}
\end{minipage}%
}%
\subfigure[Orientation error]{
\begin{minipage}[t]{0.45\linewidth}

\includegraphics[width=2.5in]{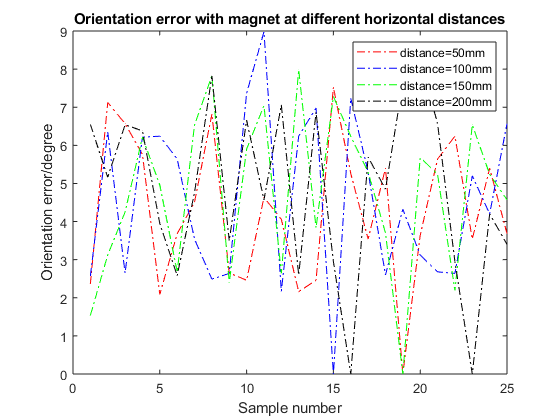}
\end{minipage}}
\caption{ The localization error and orientation error of magnet  at different horizontal distance}
\end{figure}

When the magnet is at a horizontal distance of 50 mm from the sensor array, the average positioning error at this time is 2.46 mm, of which the maximum positioning error is 5.16 mm, the minimum positioning error is 1.65 mm; the average orientation error is 4.56 degrees, of which the maximum orientation error It is 7.52 degrees, and the minimum orientation error is 2.16 degrees;
When the magnet is at a horizontal distance of 100 mm from the sensor array, the average positioning error at this time is 4.0 mm, of which the maximum positioning error is 7.68 mm, the minimum positioning error is 2.48 mm; the average orientation error is 5.12 degrees, of which the maximum orientation error It is 11.5 degrees, and the minimum orientation error is 2.17 degrees;
When the magnet is at a horizontal distance of 150 mm from the sensor array, the average positioning error at this time is 3.7 mm, of which the maximum positioning error is 6.15 mm, the minimum positioning error is 1.66 mm; the average orientation error is 5.03 degrees, of which the maximum orientation The error is 11.32 degrees, and the minimum orientation error is 1.54 degrees;
When the magnet is at a horizontal distance of 200 mm from the sensor array, the average positioning error at this time is 4.32 mm, of which the maximum positioning error is 9.87 mm, and the minimum positioning error is 2.16 mm; the average orientation error is 5.45 degrees, of which the maximum orientation The error is 12.4 degrees, and the minimum orientation error is 2.58 degrees;
It can be obtained from the experimental results that when the magnet is at a horizontal distance between 0 and 200 mm, the experimental results of positioning error and orientation error are quite different.

The experimental results show that when the magnet moves in the horizontal distance and the vertical height, the error in the horizontal direction is relatively large. We guess that the reason for this phenomenon is that when the vertical height moves, the Hall effect sensor is around the magnet, and the data can be obtained uniformly; when the magnet moves in the horizontal direction, the magnet belongs to the sensor array. A state where the unilateral magnetic field is strong and high. In order to verify this conclusion, we placed the magnet around the sensor array, as shown in Figure 10.

\subsection{Experiment-3}
Experiment with the same magnet placed in different positions. As shown in Figure 10, the permanent magnet robot is used as the source of the magnetic field. We place the magnets in different positions around the sensor array, calculate and compare the positioning accuracy. The green dots in Figure 10 represent different positions: magnet No.1, magnet No.2, magnet No.3, magnet No.4, and magnet No.5. At this time, the magnet is tested at those five different positions, and the distance between the sensor and the magnet is maintained at a position of 30 mm, that is, the distance is the vertical height of the test in this experiment.

20 sensors are used to collect data in this experiment, and the magnet is placed in different positions (magnet No.1, magnet No.2, magnet No.3, magnet No.4, and magnet No.5.) for comparison experiments. Table 1 and table 2 show the corresponding experimental results on position and orientation of the magnet, respectively. The experimental results is better when the magnet is placed around the sensor from table 1 and table 2. When the magnet is placed in the middle of the sensor(magnet No.3), that is, the sensor surrounds the magnet, the obtained position and orientation errors are relatively small.

\begin{figure}	

\center
\includegraphics[width=10 cm]{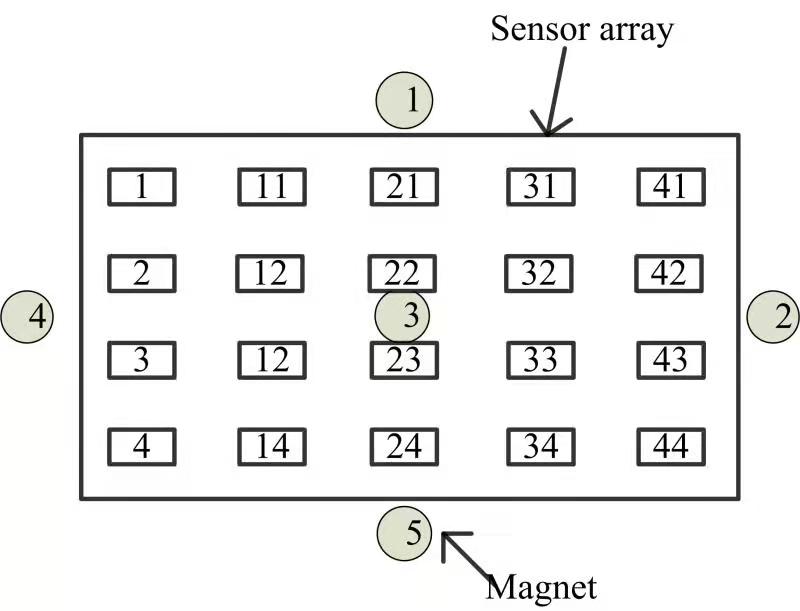}
\caption{Schematic diagram of the relative position of the permanent magnet and the sensor array.The green circle in the picture represents the position of the permanent magnet.}
\end{figure}


\begin{table}[]
\caption{the localization accuracy of the magnet at different point. }

\center
\begin{tabular}{llllll}
Position error                  &  No.1 & No.2 & No.3 & No.4 & No.5 \\
Average(mm)        & 3.17        & 2.9       & 0.47      & 2.99      & 3.06      \\
Maximum(mm)        & 4.68        & 4.01       & 2.21      & 4.18      & 4.35      \\
Minimum(mm) & 1.65        & 1.5       & 0       & 0       & 1.2      
\end{tabular}
\end{table}


\begin{table}[]
\caption{the orientation accuracy of the magnet at different point. }
\center
\begin{tabular}{llllll}
Average(degree)        & 3.7        & 3.89       & 0.92      & 3.83      & 3.87      \\
Maximum(degree)        & 7.8        & 9.65       & 2.6      & 9.27      & 7.64      \\
Minimum(degree) & 0        & 0      & 0       & 0       & 0         
\end{tabular}
\end{table}

\begin{figure}[htbp]
\centering
\subfigure[Localization error]{
\begin{minipage}[t]{0.45\linewidth}
\includegraphics[width=2.5in]{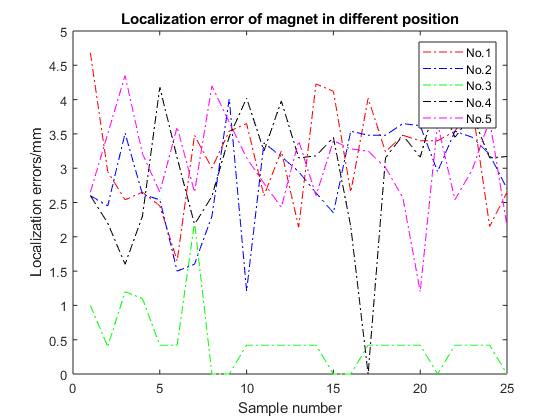}
\end{minipage}}
\subfigure[Orientation error]{
\begin{minipage}[t]{0.45\linewidth}

\includegraphics[width=2.5in]{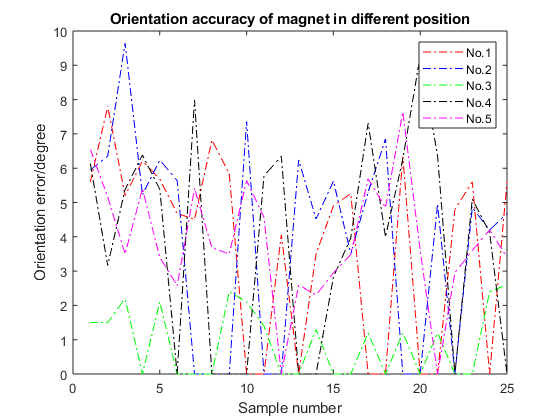}
\end{minipage}}
\caption{ The signal of magnetic field intensity when filtering is added or not.The red, yellow, and green dots in the picture represent the magnetic field intensity of the x-, y-, and z axes respectively.
}
\end{figure}

When the magnet is in the No.1 position, the average positioning error at this time is 3.17 mm, of which the maximum positioning error is 4.68 mm, and the minimum positioning error is 1.65 mm; the average orientation error is 3.7 degrees, of which the maximum orientation error is 7.8 degrees , the smallest orientation error is 0 degrees.
When the magnet is in the No.2 position, the average positioning error at this time is 2.9 mm, of which the maximum positioning error is 4.01 mm, the minimum positioning error is 1.5 mm; the average orientation error is 3.89 degrees, and the maximum orientation error is 9.65 degrees , the smallest orientation error is 0 degrees.
When the magnet is in the No.3 position, the average positioning error at this time is 0.47 mm, of which the maximum positioning error is 2.21 mm, the minimum positioning error is 0 mm; the average orientation error is 0.92 degrees, of which the maximum 
orientation error is 2.6 degrees , the smallest orientation error is 0 degrees.
When the magnet is in the No. 4 position, the average positioning error at this time is 2.99 mm, of which the maximum positioning error is 4.18 mm, the minimum positioning error is 0 mm; the average orientation error is 3.83 degrees, and the maximum orientation error is 9.27 degrees , the smallest orientation error is 0 degrees.
When the magnet is in No.5 position, the average positioning error at this time is 3.05 mm, of which the maximum positioning error is 4.35 mm, the minimum positioning error is 1.2 mm; the average orientation error is 3.87 degrees, of which the maximum orientation error is 7.64 degrees , The minimum orientation error is 0 degrees.
From the experimental results, we can see that when the sensors are evenly distributed around the magnet, the positioning accuracy is higher than that of the unilateral sensor.


\section{Conclusions}

In order to analyze the relationship between the positioning accuracy and the sensor array, we use two different layouts of Hall-effect sensors for comparison experiments. The experimental results show that when the positioning error decreases with the increase of the number of sensors,
the positioning accuracy of 20 sensors is higher than that of 16 sensors, the average positioning error is reduced by 4.02 mm. 
The magnet to be located at different vertical heights and horizontal distances from the sensor for experiments, The experimental results show that when the positioning error increases as the distance between the magnet and the sensor array increases, when the magnet is in the range of 50mm*50mm*50 mm, the average positioning error at this time is 0.762 mm, and the average orientation error is 2.1 degrees.
Magnets are placed in different positions of the sensor array for experiments, the experimental results show that when the magnets are located around the sensor array, that is, when the sensors are evenly distributed around the magnet, the error is the smallest at this time, the average positioning error at this time is 0.47 mm, and the average orientation error is 0.92 degrees.

The experimental results in this article are all obtained in a static magnetic field environment. We will study the positioning in a dynamic magnetic field in the next future.


\vspace{6pt} 

\footnotesize 

\printbibliography 
\end{document}